\documentclass[aps,pre,showpacs,twocolumn]{revtex4}
\usepackage{amssymb}
\usepackage{graphicx}
\usepackage{colordvi}
\usepackage{color}
\usepackage{dcolumn,amsmath,amsthm,amscd,amsfonts,amssymb,epsfig,graphics,graphicx,eucal}
\usepackage[dvipsnames]{xcolor}
\pdfoutput=1
\newcommand{\PT}{{\cal PT}}

\begin{document}

\title{Localized modes in   $\chi^{(2)}$ media with $\PT$-symmetric localized potential}

\author{F. C. Moreira$^{1,2}$, F. Kh. Abdullaev$^{3}$, V. V. Konotop$^{2}$ and A. V. Yulin$^2$ }

\affiliation{
$^1$Universidade Federal de Alagoas, Campus A. C. Sim\~oes - Av. Lourival Melo Mota, s/n, Cidade Universit\'aria, Macei\'o - AL 57072-900, Brazil
\\
$^2$Centro de F\'isica Te\'orica e Computacional and Departamento de F\'isica, Faculdade de Ci\^encias, Universidade de Lisboa, Avenida Professor Gama Pinto 2, Lisboa 1649-003, Portugal
\\
$^3$Instituto de Fisica Te\'orica, Universidade Estadual Paulista, R. Dr. Bento Teobaldo Ferraz, 271 Barra Funda, S\~ao Paulo, CEP 01140-070, Brasil}

\date{\today}

\begin{abstract}
We study the existence and stability of solitons in the quadratic nonlinear media with spatially localized ${\cal PT}$-symmetric modulation of the linear
refractive index. Families of stable one and two hump  solitons are found. The properties of nonlinear modes bifurcating  from a linear limit of small fundamental harmonic field are investigated.
It is shown that the fundamental branch, bifurcating from the linear mode of the fundamental harmonic is limited in power. The power maximum  decreases with the strength of the imaginary part of the refractive index. The modes bifurcating from the linear mode of the second harmonic can exist even above ${\cal PT}$ symmetry breaking threshold.
We found that the fundamental branch bifurcating from the linear limit can undergo a secondary bifurcation colliding with a branch of two-hump soliton solutions. The stability intervals for different values of the propagation constant and gain/loss gradient are obtained. The examples of dynamics and excitations of solitons obtained by numerical simulations are also given.
\end{abstract}
\maketitle



\section{Introduction}

Non-Hermitian Hamiltonians satisfying
the parity-time ($\PT$) symmetry can have real eigenvalues~\cite{Bender1}.
While originally these ideas were developed in the quantum-mechanical context, it became soon clear that new broad applications they can find in optics. The first suggestions of  the optical analogues of the $\PT$  symmetric Hamiltonians was proposed in~\cite{Muga} and was based on a linear planar waveguide structure.
$\PT$ symmetry effect predictions are confirmed in experiments with light propagation in couplers with gain and loss~\cite{Exp1}.
Later on it was suggested to explore nonlinear optical media obeying  $\PT$-symmetry, and in particular it was shown the possibility of soliton propagation in such media~\cite{Christodoulides1}.
The $\PT$-symmetry was modeled by the refractive index having symmetric real part  and anti-symmetric imaginary part.  A particularly interesting realization of a $\PT$ symmetric modulation of the refractive index is when losses and gain are localized in space, giving rise to a $\PT$ symmetric localized impurity. Such impurities allow for existence of localized (defect) modes. In the linear theory such modes were studied for exactly integrable models in~\cite{Znojil,Ahmed}, and their nonlinear extension was reported in~\cite{Christodoulides1}. Solitons supported by other $\PT$ symmetric defects were also reported for focusing~\cite{Gaussian} and defocusing~\cite{defoc} media. Linear scattering by a $\PT$-symmetric inhomogeneity and emerging of the related spectral singularities was described in~\cite{Mostafa_scattering}. The effect of two and various randomly distributed $\PT$ symmetric impurities on the lattice dynamics was addressed in~\cite{Kottos}. Switching of solitons in a unidirectional coupler using $\PT$-symmetric defects was suggested in~\cite{AKOS}. Nonlinear modes in even more sophisticated, double well $\PT$-symmetric potentials were studied recently~\cite{double_well}.

All of the works mentioned above and devoted to nonlinear modes dealt with the PT symmetric media possessing Kerr nonlinearity (see the list of recent works on solitons in~\cite{PT_sol}). It is a natural further step to address a possibility of existence of defect modes and their stability in another class of widely used optical media, which is are the $\chi^{(2)}$ materials. Solitons in quadratic nonlinear media with conservative defects were investigated in \cite{Clausen1,Clausen2}, where it was shown that solutions are dynamically stable in the case of attractive impurities. In the present paper we study the existence of solitons in the media with quadratic nonlinearity and localized $\PT$-symmetric potentials.

The paper is organized as follows.
In Section II the model and statement of the problem are formulated. The properties of localized modes for different ratios between fundamental and second harmonics are studied in Sections III-V. The stability and dynamics of localized solutions  are considered in Section VI.

\section{Statement of the problem}

We consider the system
\begin{subequations}
\label{final}
\begin{align}
    i\frac{\partial u_{1}}{\partial\zeta}+\frac{\partial^{2}u_{1}}%
{\partial\xi^{2}}+V\left( \frac{1}{\cosh^2 \xi}  +i\alpha
\frac{\sinh  \xi}{\cosh^2  \xi}  \right)
u_{1}
+2\overline{u}_{1}u_{2}=0,
\label{final1}
\\
i\frac{\partial u_{2}}{\partial\zeta} +\frac{1}{2}\frac{\partial^{2}u_{2}%
}{\partial\xi^{2}}+ 2\left(  \frac{V}{\cosh^2 \xi}   +q\right)
u_{2}+u_{1}^{2}=0  \label{final2}%
\end{align}
\end{subequations}
describing spatial second-harmonic generation in a $\chi^{(2)}$ material with localized modulation of the refractive index, $u_1$ and $u_2$ being the dimensionless fields of the first and second harmonics, $\xi$ and $\zeta$ are the dimensionless transverse and propagation coordinates, scaled to the characteristic size of the modulation of the refractive index which is characterized by the amplitude $V$. The mismatch in the propagation constants of field components is described by $q$.  We notice that experimentally the introduced model can describe a medium with active dopants, typically having rather narrow spectral resonances, i.e. affecting only a limited range of frequencies. In particular, such impurities can induce gain and dissipation, whose strengths is characterized by $\alpha$, only for one of the field component, which in our case is the FF.  

 {Before into the detail study of the system (\ref{final})  we note, that in the standard way  solitonic solutions can be found in the analytical form in the limit of large mismatch parameter $-q \gg 1$, when  $u_2 \approx -u_1^2/(2q)$ and  the system  (\ref{final}) reduces to the NLS equation with PT-symmetric potential for the fundamental harmonic $u_1$ 
\begin{equation}
i\frac{\partial u_1}{\partial\zeta} + \frac{\partial^2 u_1}{\partial \xi^2} + V(\frac{1}{\cosh^2 \xi} +i\alpha
\frac{\sinh  \xi}{\cosh^2  \xi}) - \frac{1}{q}|u_1|^2 u_1 =0.
\end{equation}
For $q< 0$ the bright soliton solution has the form~\cite{Christodoulides1}
\begin{eqnarray}
u_1 = \sqrt{|q|A}\mbox{sech}(\xi)\exp[i\frac{\alpha V}{3}\tan^{-1}(\sinh(\xi)) + i\zeta], \nonumber\\
A = (2-V+\frac{\alpha^2 V^2}{9})
\end{eqnarray}}

We are interested in the localized solutions
\begin{equation}
u_{n}\left(  \xi,\zeta\right)  =w_{n}\left(  \xi\right)  e^{in b\zeta},%
\quad n=1,2 ,\label{statio}%
\end{equation}
where $b$ is the propagation constant of the first harmonic and $w_{1,2}$  solves the system
\begin{subequations}
\label{stat}%
\begin{equation}
\frac{d^{2}w_{1}}{d\xi^{2}}+\left[  V\left(  \frac{1}{\cosh^2 \xi}  +i\alpha
\frac{\sinh  \xi}{\cosh^2  \xi}\right)    -b\right]  w_{1}+
2\overline{w}_{1}w_{2} =0,\label{stat1}
\end{equation}
\begin{equation}
\frac{1}{2}\frac{d^{2}w_{2}}{d\xi^{2}}+2\left(  V \frac{1}{\cosh^2 \xi} +q-b\right)w_{2}+w_{1}^{2} =0. \label{stat2}%
\end{equation}
\end{subequations}
subject to the zero boundary conditions $ w_{1,2}\left(  \xi\right)\to 0 $ as $|\xi|\to \infty$.

We restrict our consideration mainly to solutions bifurcating from the linear limit, which is understood as a limit where at least one of the harmonics vanishes. It follows from (\ref{stat}), that so defined linear limit does not necessarily implies that both amplitudes $w_1$ and $w_2$ are infinitely small. Indeed it is sufficient to require that only the amplitude of the fundamental harmonic $w_1$ is infinitely small to consider the equations (\ref{stat}) in the linear limit. That is why one can distinguish three different bifurcations of the fundamental soliton solution from the linear limit:

(i) The amplitude of the second harmonic is of the order of the squared amplitude of the first harmonic, i.e. is negligible compared to the amplitude of the first harmonic
\begin{eqnarray}
\label{case1}
w_2=O( w_1^2) \qquad w_1\to 0
\end{eqnarray}

(ii) The second harmonic is finite in the limit of negligible first harmonic
\begin{eqnarray}
\label{case2}
  w_2=O( 1) \qquad w_1\to 0
\end{eqnarray}

(iii) Both harmonics are of the same order
\begin{eqnarray}
\label{case3}
w_2= O (w_1) \qquad w_1\to 0
\end{eqnarray}

As it follows from (\ref{case1}) and (\ref{case3}) the  maximal intensities of both $w_1$ and $w_2$  go to zero at the bifurcation point for cases (i) and (iii). However in case (ii) the maximum of the absolute value of field $w_2$ goes to a finite value when the solution approaches the bifurcation point. Which of the cases is realized, depends on the parameters of the system, and in particular on the mismatch $q$. Below we consider these three cases separately.

\section{Modes with negligible second harmonic in the linear limit.}

Let us start with the conditions necessary for (\ref{case1}) to occur. In this limit  the nonlinear term in (\ref{stat1}) can be neglected and in the leading order we have the eigenvalue problem
\begin{subequations}
\begin{equation}
L_{1,\alpha}w_{1l}=bw_{1l},
\end{equation}
\begin{equation}
L_{1,\alpha}=\frac{d^{2}}{d\xi^{2}}+ V\left(\frac{1}{\cosh^2 \xi}  +i\alpha\frac{\sinh  \xi}{\cosh^2  \xi}\right).
\end{equation}
\label{lin1}
\end{subequations}
Eq.~(\ref{lin1}) has been studied before. Therefore below we only briefly outline the features necessary for our analysis, referring to~\cite{Znojil,Ahmed} for more details.
\begin{figure}[h]
\begin{center}
\scalebox{0.67} {\includegraphics{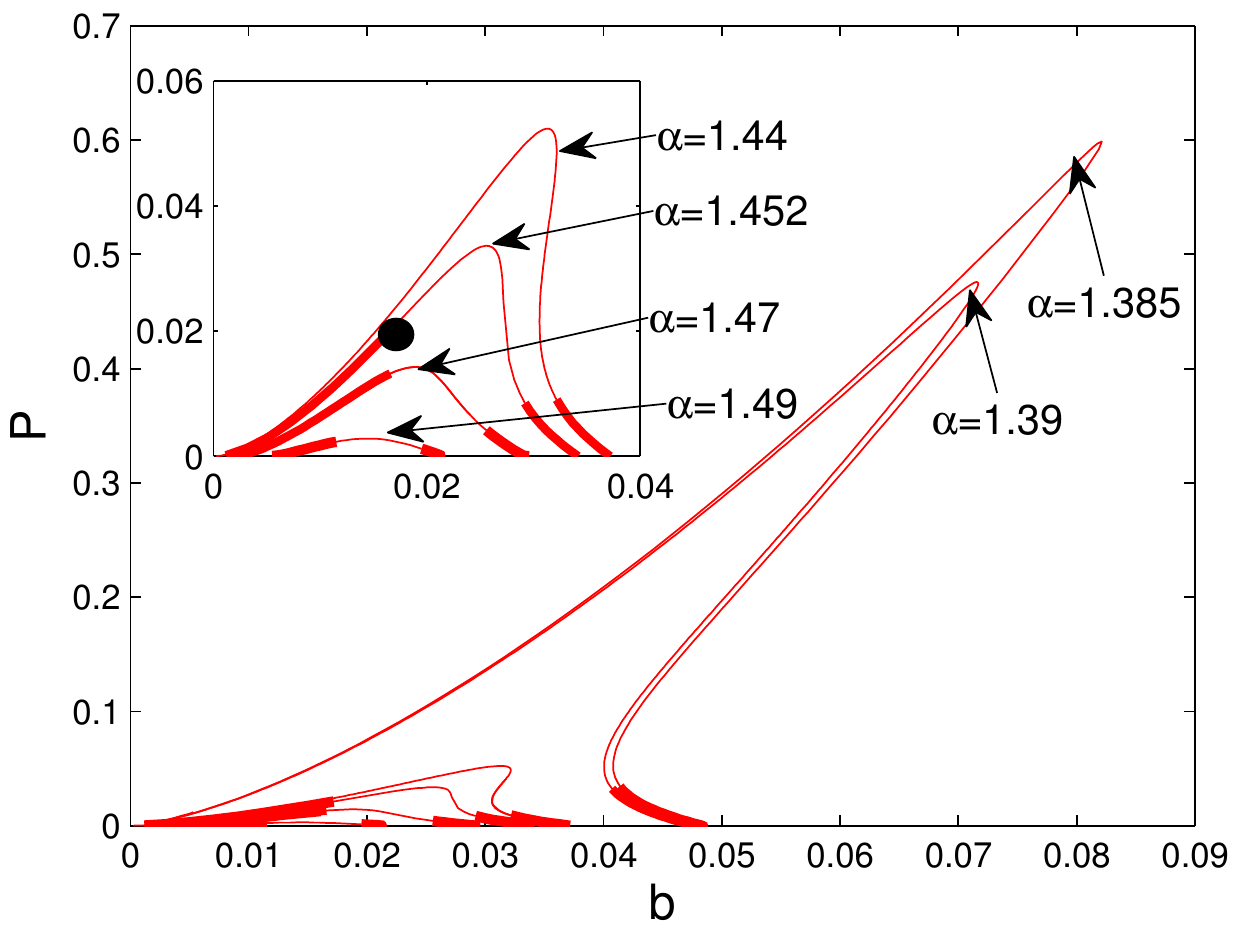}}
\end{center}
\caption{(Color online) Families of the solutions bifurcating from $b_{1,0}=\eta_1^2$ in the case (i) for several values of $\alpha$. Thick (thin) lines represent stable (unstable) solutions.  Insertion shows the branches for larger values of $\alpha$. The black circles represent solutions studied in text. The parameters of the structure are $V=1/2$ and $q=0$ and $\alpha_{cr}=1.5$.}%
\label{fig:branchsFF}%
\end{figure}
For $V>0$ Eq. (\ref{lin1}) possesses localized solutions.
When $\alpha$ is below the $\PT$-symmetry breaking threshold
\begin{equation}
\label{PT-breaking}
\alpha < \alpha_{cr}=1+\frac{1}{4V},
\end{equation}
the spectrum of Eq.~(\ref{lin1}) has discrete real eigenvalues given by~\cite{Midya}
\begin{equation}
b_{1,n}=(n-\eta_1)^2, \quad n=0,1,...<\eta_1,
\label{realeig1}
\end{equation}
where
\begin{equation}
\eta_1=\frac 12 \left(\sqrt{V(\alpha_{cr}-\alpha)}+\sqrt{V(\alpha+\alpha_{cr})} -1\right).
\label{n1}
\end{equation}
Above the symmetry breaking point ($\alpha>\alpha_{cr}$) the eigenvalues of the bound states are complex valued. We notice here that no fundamental branch satisfying condition (\ref{case1}) with $\alpha>\alpha_{cr}$ was found.
\begin{figure}[h]
\begin{center}
\scalebox{0.67} {\includegraphics{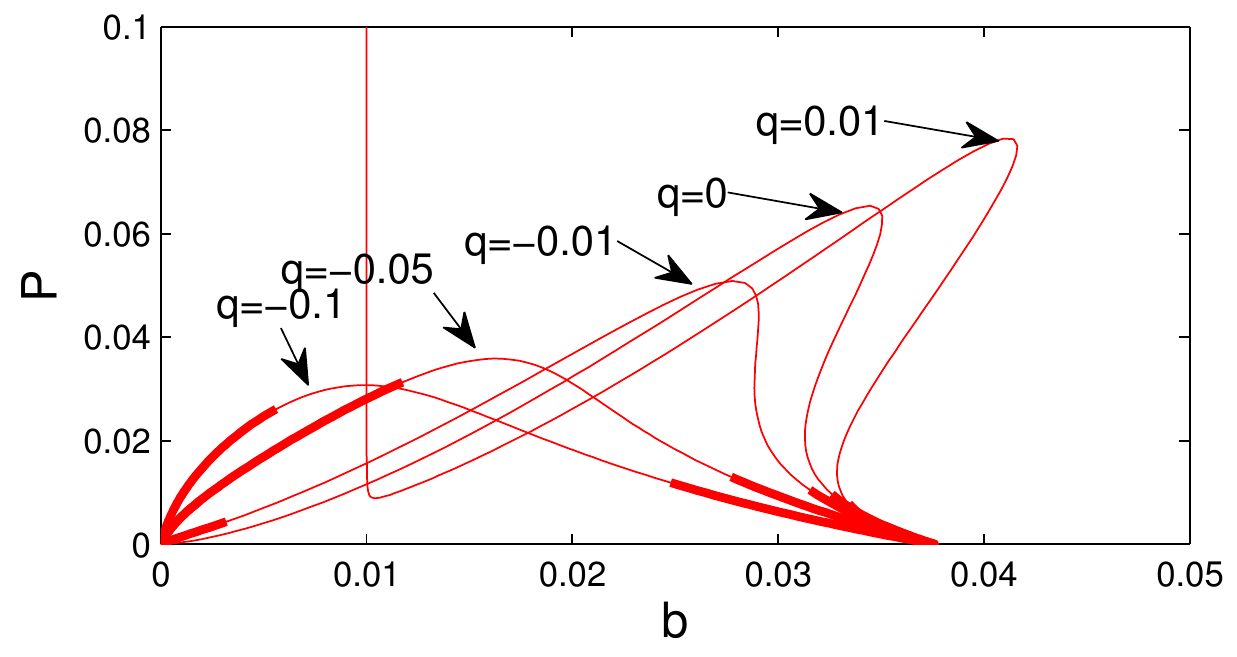}}
\end{center}
\caption{(Color online) Families of the solutions bifurcating from $b_{1,0}=\eta_1^2$ and corresponding to the case (i) for several values of $q$. Thick (thin) lines  represent stable (unstable) solutions. The parameters of the structure are $V=1/2$ and $\alpha=1.44$.}%
\label{fig:branchesalpha144}%
\end{figure}
Therefore from now on  we concentrate only on the results for $\PT$-symmetry preserving case (\ref{PT-breaking}). Moreover,
 our consideration will be limited  to nonlinear modes that bifurcate from the ground state of defect potential in Eq.~(\ref{lin1}), i.e. $n=0$. 
 The respective eigenstate of the linear problem (\ref{lin1}) reads~\cite{Znojil,Ahmed}
\begin{equation}
w_{1l}(\xi)=W_1\text{sech}^{\eta_1}(\xi)\exp\left[\frac{i}{2}\Theta\tan^{-1}(\sinh\xi)\right]
\label{w1lpt}
\end{equation}
where $W_{1}$ is a  constant and
\begin{equation}
\Theta=\sqrt{V(\alpha_{cr}-\alpha)}-\sqrt{V(\alpha+\alpha_{cr})}.
\end{equation}

Passing to (\ref{stat2}), in the small amplitude limit one can look for a solution with $w_1\approx w_{1l}$, which plays the role of the inhomogeneous term in the linear equation for $w_2$.
However, to obtain the complete families of solutions one has to consider both Eqs.~(\ref{stat}).  We did this using relaxation Newton-Raphson method using the described linear solutions as the initial ansatz.

Fig.~\ref{fig:branchsFF}  
shows the dependence of the total power
\begin{equation}
P=P_1+P_2, \qquad P_{n}=n\int_{-\infty}^{\infty}|w_{n}|^{2}d\xi
\label{power}
\end{equation}
on the propagation constant with several values of $\alpha$.
We observe that the fundamental branches shown in Fig.~\ref{fig:branchsFF} have a maximal in power $P_M$, $P\leq P_M$. The value of $P_M$ is decreasing as $\alpha$ increases. We also observed that $P_M\to 0$ as $\alpha\to \alpha_{cr}$, i.e. the fundamental branch disappears.
\begin{figure}[h]
\begin{center}
\scalebox{0.5} {\includegraphics{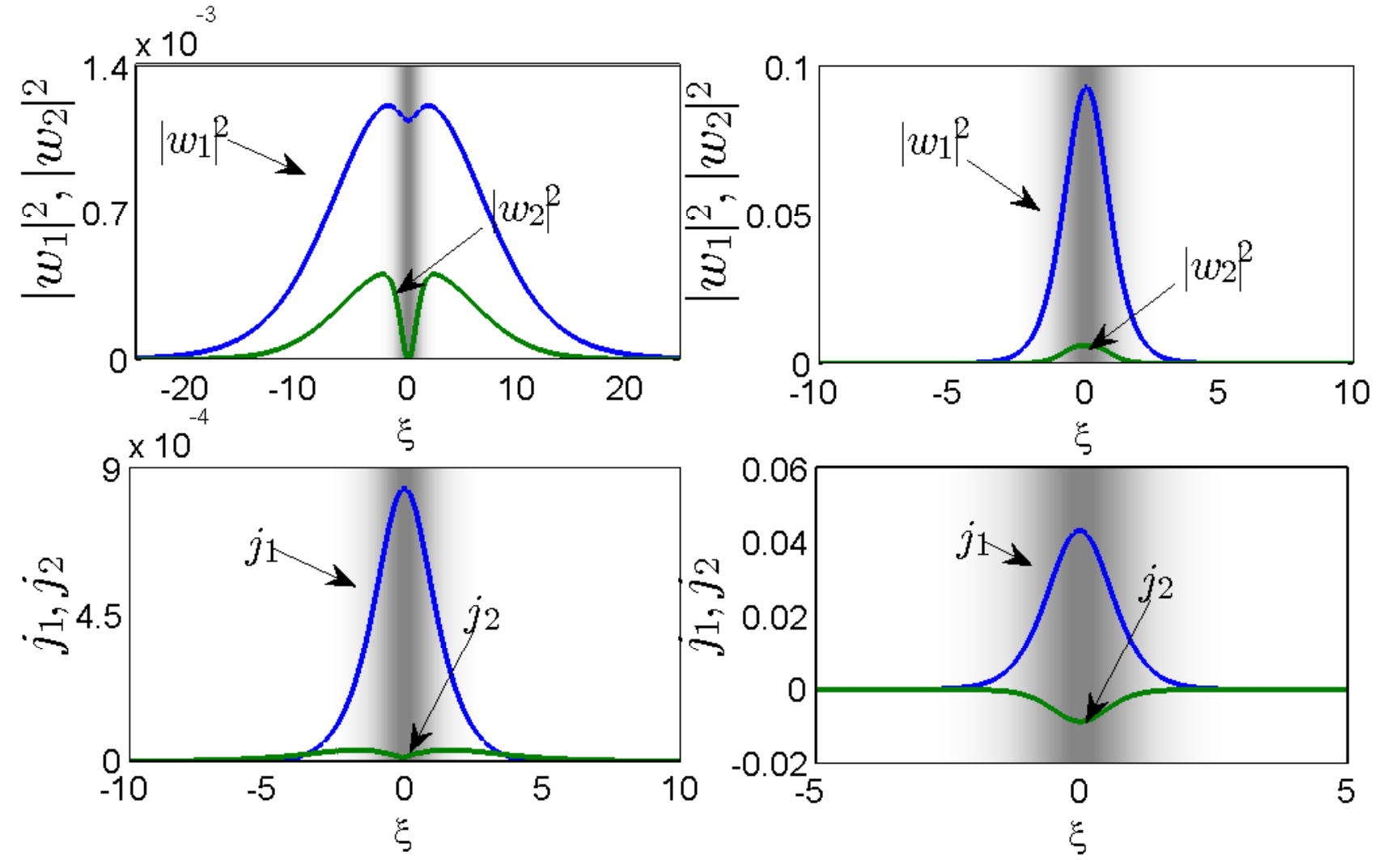}}
\end{center}
\caption{(Color online) Spatial distributions of the intensities $|w_n|^2$ (upper panels) and  the currents $j_n$ (lower panels). The left panels corresponds to a stable solution with $b=0.024$, as marked by a black circle in Fig. \ref{fig:branchsFF}, pertaining to the fundamental branch that bifurcates from $b_{1,0}=0.034$, where $V=1/2$, $q=0$ and $\alpha=1.452$. The right panels corresponds to a stable solution with $b=0.82$, pertaining to the fundamental branch that bifurcates from $b_{1,0}=0.92$, where $V=2$, $q=0$ and $\alpha=0.5$. Shadowed domains show the localized impurity (darker areas represent higher values of its real part, $V\text{sech}^{2}\xi$.}
\label{fig:solutionsalpha144}%
\end{figure}
For a given $\alpha$, the position of $P_M$ in respect to $b$  approaches $b=0$ as $q$ decreases, as one can see in Fig.~\ref{fig:branchesalpha144}. The position of $P_M$ moves to the right in the case $q>0$. One can also see in Fig.~\ref{fig:branchesalpha144} that in this case the branch ends in $b=-q$ because the mismatch shifts the position of the continuum spectrum of the linear part of (\ref{stat2}) and $w_2$ becomes delocalized. There is no low amplitude linear limit for $w_{1}$ in this case.  

In Fig.~\ref{fig:solutionsalpha144} we show the typical distributions of the fields $w_n$ and the real-valued currents $j_n$ defined as ($n=1,2$)
\begin{equation}
j_{n}(\xi)=|w_{n}|^2\frac{d \theta_{n}}{d\xi},
\qquad
\theta_{n}(\xi)=\text{arg} w_n(\xi).
\end{equation}
By construction $|w_n|^2$ and $j_n$ are even functions.
The effective width of the intensities of modes may significantly exceed the size of the impurity, particularly in the modes closer to the edge of the continuous spectrum (the left panels of Fig.~\ref{fig:solutionsalpha144}). 

We note that in the left panels of Fig.~\ref{fig:solutionsalpha144} the solution has a relatively small amplitude. This is a peculiarity of the chosen strength of the potential (it was $V=1/2$, solitons with larger amplitudes were found to be unstable). While this potential (ensuring the existence of only one linear defect level in the localized potential) is used below along the text, in the right panels of Fig.~\ref{fig:solutionsalpha144} we show a higher amplitude soliton for the potential well having the width $V=2$ (and the parameters $\alpha=0.5$ and $q=0$).

\section{Nonlinear modes without linear limit.}

In this section we investigate the case when the second harmonic remains finite at $w_1 \rightarrow 0$. Then one can neglect the nonlinear term $w_{1}^{2}$ in (\ref{stat2}) reducing it to the well known linear eigenvalue problem (see e.g.~\cite{Landau})
\begin{subequations}
\begin{equation}
L_{2}w_{2l}=2bw_{2l},
\end{equation}
\begin{equation}
L_{2}=\frac{1}{2}\frac{d^{2}}{d\xi^{2}}+2\left(  V\frac{1}{\cosh^2 \xi}+q\right),
\end{equation}
\label{lin2}
\end{subequations}
whose eigenvalues are 
\begin{equation}
b_{2,n}=\frac{\left(n-\eta_2\right)^2}{4}+q, \quad n<0,1,...,\eta_2=\sqrt{\frac 14+4V}-\frac 12
\label{realeig2}
\end{equation}
Here we again consider the case when there is only one localized mode. This leads to the requirement that $\eta_2\leq 1$ and consequently $V\leq 1/2$.
 The corresponding eigenfunction of $b_{2,0}$ reads
\begin{equation}
w_{2,0}(\xi)=W_2\text{sech}^{\eta_2}\xi.
\label{w2l}
\end{equation}
where $W_2$ is some constant which must be determined.
\begin{figure}[h]
\begin{center}
\scalebox{0.51} {\includegraphics{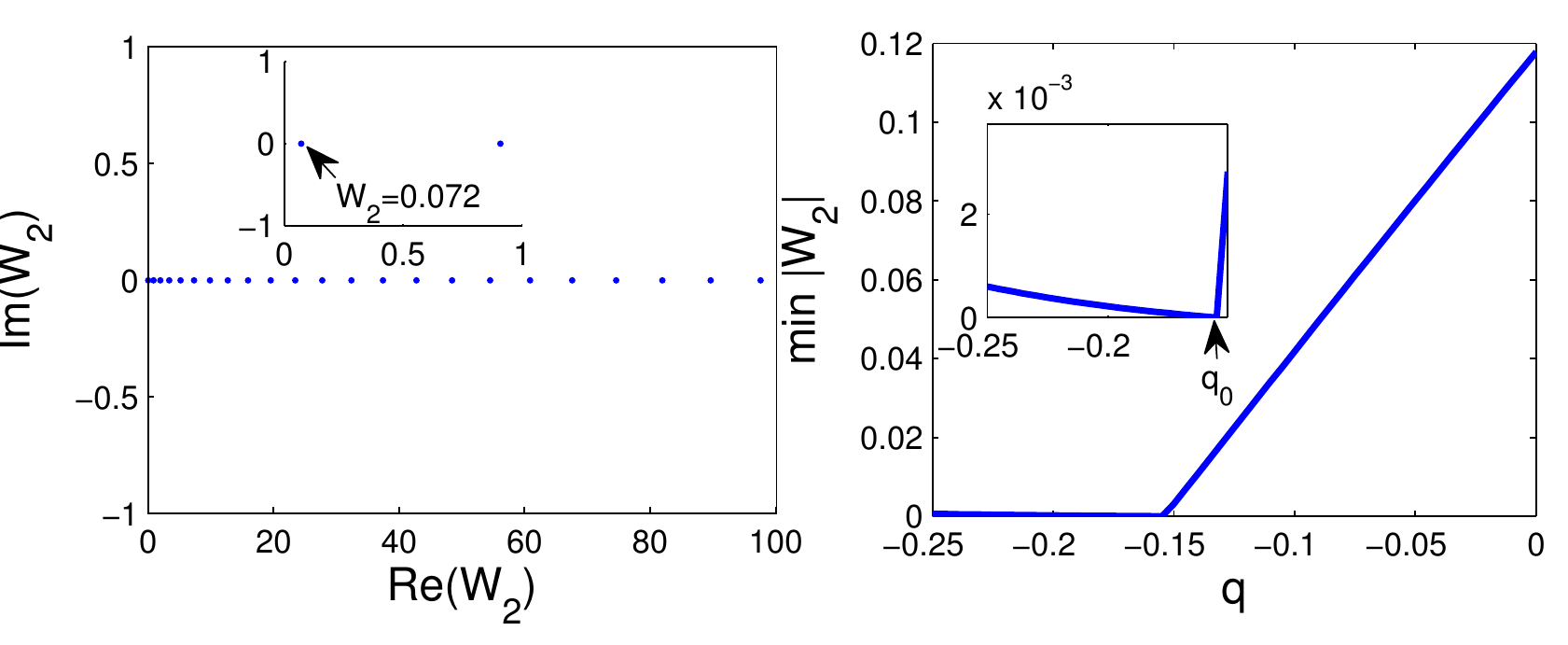}}
\end{center}
\caption{(Color online) Left panel: The eigenvalues $W_2$ of (\ref{w10}). The parameters of the structure are $V=1/2$, $\alpha=1.4$ and $q=0$. Insertion shows the first few eigenvalues in detail. Right panel: The lowest $\left|W_2\right|$ of (\ref{w10}) as a function of the mismatch $q$ of the lowest $P$ branch. The insertion shows how the minimum value of $W_2$ reaches zero at $q=q_{0}$.}%
\label{fig:parameterW2}%
\end{figure}
This can be done from the condition that the propagation constants in the equations (\ref{lin2}) and (\ref{stat1}) are the same. In the vicinity of the bifurcation point one can approximate $w_2\approx w_{2l}$, i.e., Eq.~(\ref{stat1}) can be approximated by the following linear system
\begin{subequations}
\begin{equation}
L\left(
\begin{array}[c]{c}%
\bar{w}_1\\ w_1
\end{array}
\right)  =W_2
\left(
\begin{array}[c]{c}%
\bar{w}_1\\ w_1
\end{array} \right),
\label{w10m}
\end{equation}
\begin{equation}
L=-\frac 12 \cosh^{\eta_2}(\xi)\left(
\begin{array}[c]{cc}%
0& L_{1}-b_{2,0}\\
\bar{L}_{1}-b_{2,0}&0
\end{array}
\right).
\end{equation}
\label{w10}
\end{subequations}
\begin{figure}[h]
\begin{center}
\scalebox{0.66} {\includegraphics{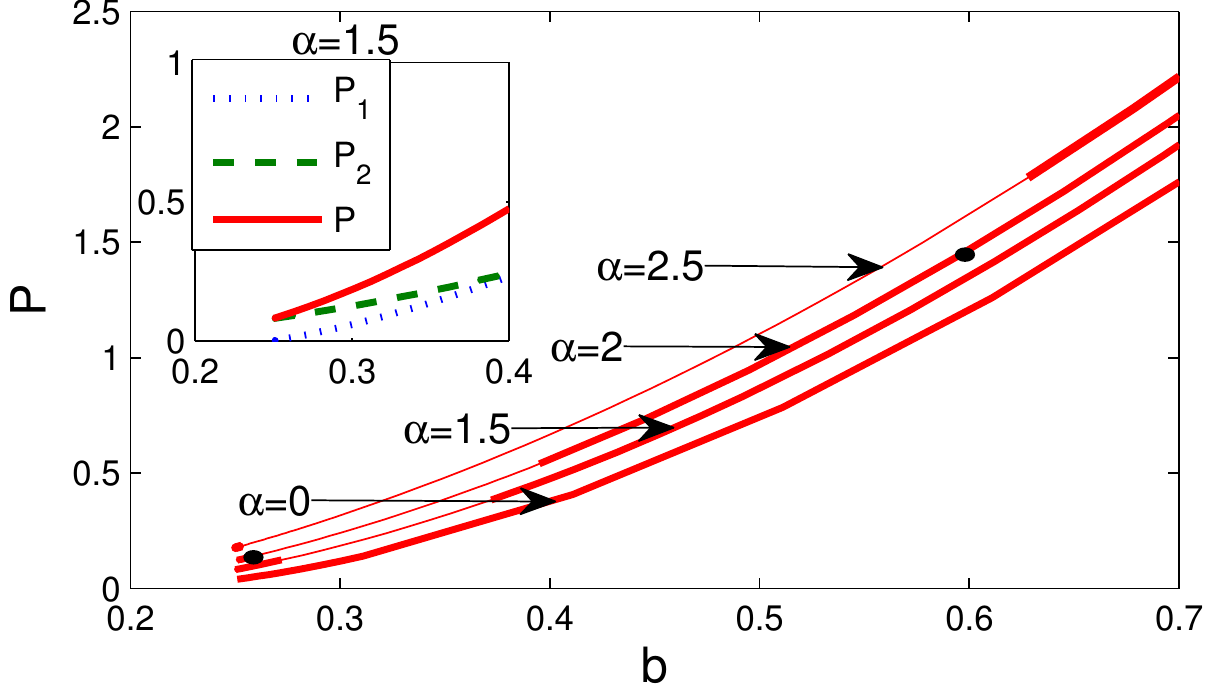}}
\end{center}
\caption{(Color online) The power diagrams of the fundamental branches bifurcating from the linear mode $b_{2,0}=0.25$ for several values of $\alpha$. Thick (thin) lines  represent stable (unstable) solutions. Insertion shows in detail that $P_1$ goes to zero in the vicinity of $b_{2,0}$ while $P_2$ remains finite. Filled circles represents two stable solutions shown below in Fig.~\ref{fig:solb06SH}. The parameters of the structure are $V=1/2$ and $q=0$.}%
\label{fig:branchSH}%
\end{figure}
Let us note that (\ref{w10m}) is an eigenvalue equation and so the allowed values of $W_2$ are simply the eigenvalues. Next we define a bra-- and  ket-- vectors:   $\langle \psi|\equiv( \psi  (\xi),\bar \psi (\xi))$ and $|\psi\rangle \equiv(\bar{\psi} (\xi),\psi (\xi))^{T}$ ($T$ stays for the transpose matrix), where $\psi (\xi)\to 0$ at $\xi\to\pm\infty$, and verify that the operator $L$ is Hermitian with respect to the weighted inner product:
\begin{equation}
\langle\psi_1|\psi_2\rangle=\int_{-\infty}^{\infty}\text{sech}^{\eta_{2}}(\xi)\left[ {\psi_1}(\xi)\bar{\psi}_2(\xi)+\bar{\psi}_1(\xi)\psi_2(\xi)\right]d\xi.
\label{dpro}
\end{equation}
The  Hermiticity of $L$ ensures the reality of the admissible $W_2$. We also note that if $\left(\bar{w}_1(\xi),w_{1}(\xi)\right)^{T}$ is a solution of (\ref{w10}) with eigenvalue $W_2$, then $-W_2$ is also an eigenvalue with eigenfunction $\left(-i\bar{w}_1(\xi),iw_{1}(\xi)\right)^{T}$. This allows us to restrict the consideration to $W_2>0$.

We investigated the system (\ref{w10}) numerically and found that there is an infinite number of discrete eigenvalues $W_{2}$. The ones having the smalest absolute values are shown in Fig.~\ref{fig:parameterW2}. 
 \begin{figure}[h]
\begin{center}
\scalebox{0.5} {\includegraphics{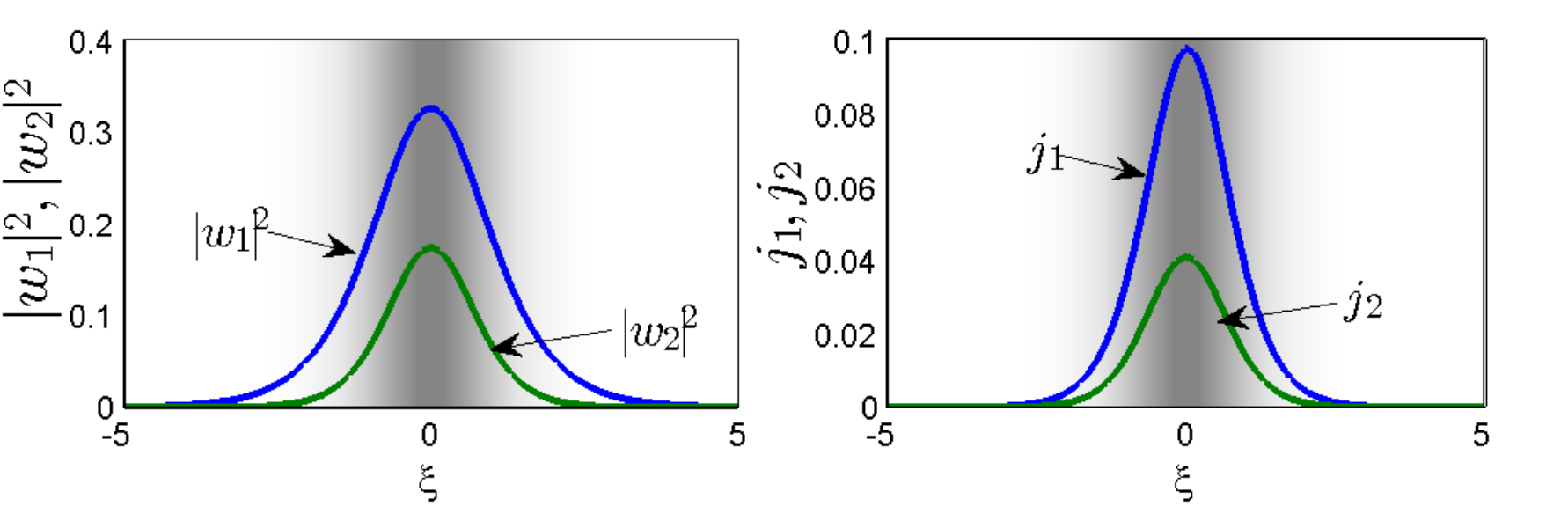}}
\end{center}
\caption{(Color online) An example of a stable solution of case (ii) with $b=0.6$ and $\alpha=2>\alpha_{cr}$, marked with the filled circle in Fig.~\ref{fig:branchSH}.  
Shadowed domains show the localized impurity (darker areas representing higher values of the real part of the localized potential). The parameters of the structure are $V=1/2$ and $q=0$.}%
\label{fig:solb06SH}%
\end{figure}
The amplitude of the second harmonic $W_2$ depends on $b=b_{2,0}$ (See Eq.(\ref{w10m})). In Fig.~\ref{fig:parameterW2}, we show the dependence of $W_2$ on $q$ corresponding to the lowest $|W_2|$ branch. The special case, when for a certain value of $q$ the amplitude of the second harmonic $W_2\to 0$, happens when $b_{1,0}=b_{2,0}$. This leads to
\begin{align}
\label{cond_q}
q=q_{0}, \qquad	q_{0}=\eta_1^2-\frac 14 \eta_2^2,
\end{align}
which is precisely the case however (\ref{case3}); we consider it in the next section.

We numerically studied the existence of bifurcations satisfying (\ref{case2}) in Fig.~\ref{fig:branchSH}. It is possible to see in the insertion of Fig.~\ref{fig:branchSH} that at the bifurcation point the branches satisfy $P_1(b_{2,0})=0$ and consequently $P(b_{2,0})=P_{2}(b_{2,0})$. A simple integration in (\ref{power})  after the substitution $w_2=w_{2l}$,  reveals that for each $\alpha$ when $V=1/2$ and $\eta_{2}=1$ (the case of Fig.~\ref{fig:branchSH}) one has $P(b_{2,0})=4W_{2}^{2}$.
 
It can be seen in Fig.~\ref{fig:branchSH} for the power diagrams of the fundamental branches, where stable solutions exist above the $\PT$- symmetry breaking point.   The existence of stable nonlinear modes even when the spectrum of the linear system is not purely real has been earlier reported in \cite{Zezyulin} (see also recent work~\cite{Tsoy}.)
 
In Fig.~\ref{fig:solb06SH} there is an example of a mode in the case (ii).

\section{Bifurcation of the nonlinear modes from the linear spectrum.}

Now we consider the case (iii) for which the relation (\ref{case3}) holds.\begin{figure}[h]
\begin{center}
\scalebox{0.65} {\includegraphics{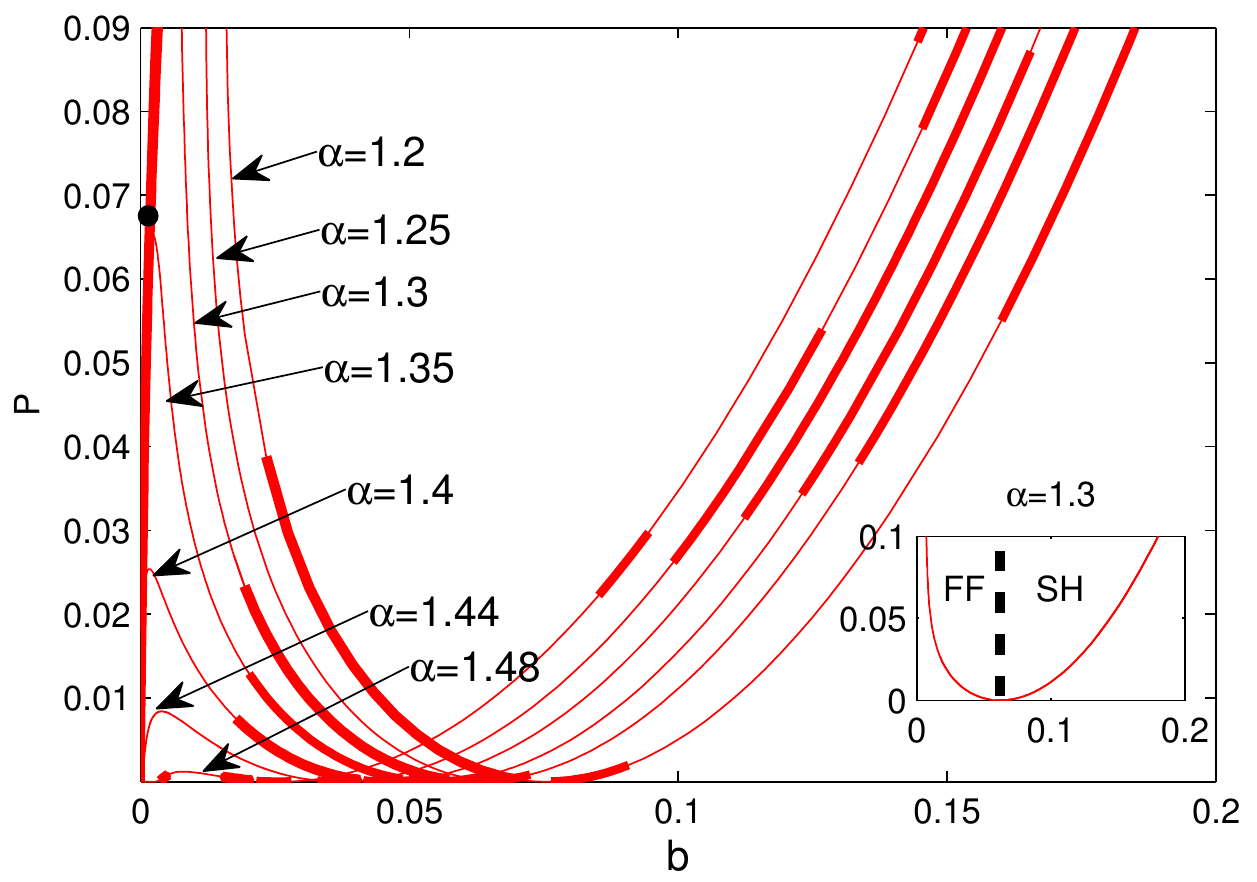}}
\end{center}
\caption{(Color online) Several fundamental branches for the case (iii) with different values of $\alpha$. Note that the branch disappear when $\alpha\to\alpha_{cr}=3/2$. Stable (unstable) solutions are represented by thick(thin) lines. Insertion shows the regions of bifurcations from $b_{1,0}$ (FF) and $b_{2,0}$ separated by a vertical dashed line. The parameters of the structure are $V=1/2$ and $q=q_0$($\alpha_{cr}=1.5$).}%
\label{fig:V05qqbranch}%
\end{figure}
Previously we have shown that if the bifurcation point is at the same time an eigenvalue of (\ref{lin1}) and (\ref{lin2}), i.e., $b_{1,0}= b_{2,0}$, then the mismatch must have the special value $q=q_0$. We also have seen that in this case $W_2=0$. As a direct consequence, (\ref{w10}) reduces to (\ref{lin1}) and not only $w_2\to w_{2l}$ but also the FF satisfy $w_{1}\to w_{1l}$ at the bifurcation point.

 The Fig.~\ref{fig:V05qqbranch} shows power diagrams of solutions satisfying (\ref{case3}) for several values of $\alpha$.
  \begin{figure}[h]
\begin{center}
\scalebox{0.65} {\includegraphics{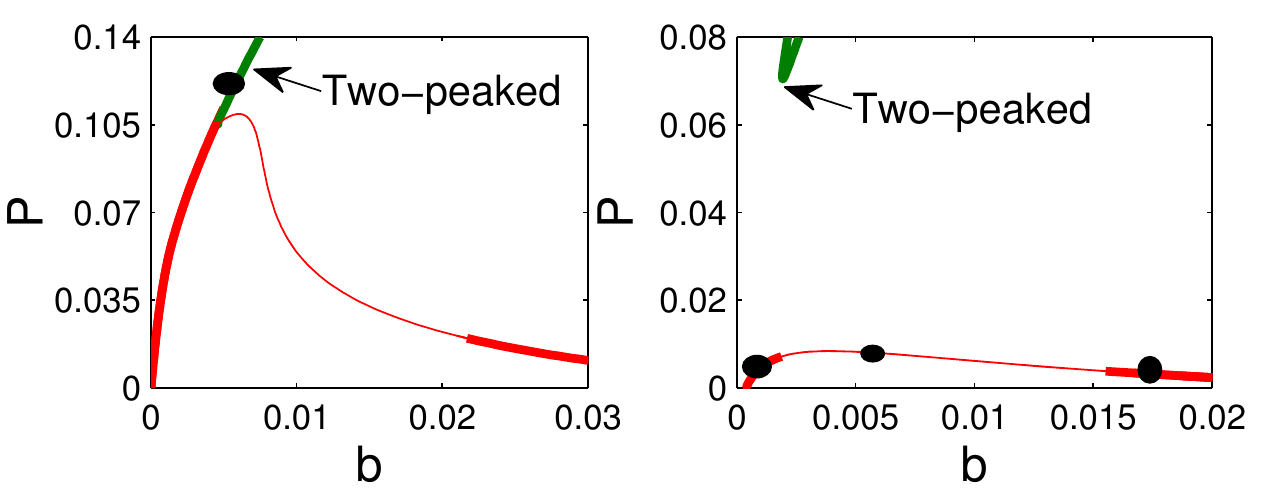}}
\end{center}
\caption{(Color online) Left panel: Shows the fundamental branch (line) of case (iii) with $\alpha=1.3$ near $b=0$ and the merged two-peaked branch. Thick lines (lines) are stable (unstable) solutions. Right panel: Shows a fundamental branch and a two-peaked branch with $\alpha=1.44$ near $b=0$. Black circles are solutions represented in Fig.~\ref{fig:twopeak} and in Fig.\ref{fig:dynstabSIMUL}. The parameters of the structure are $V=1/2$ and $q=q_0$.}  \label{fig:simb0}
\end{figure}
\begin{figure}[h]
\begin{center}
\scalebox{0.5} {\includegraphics{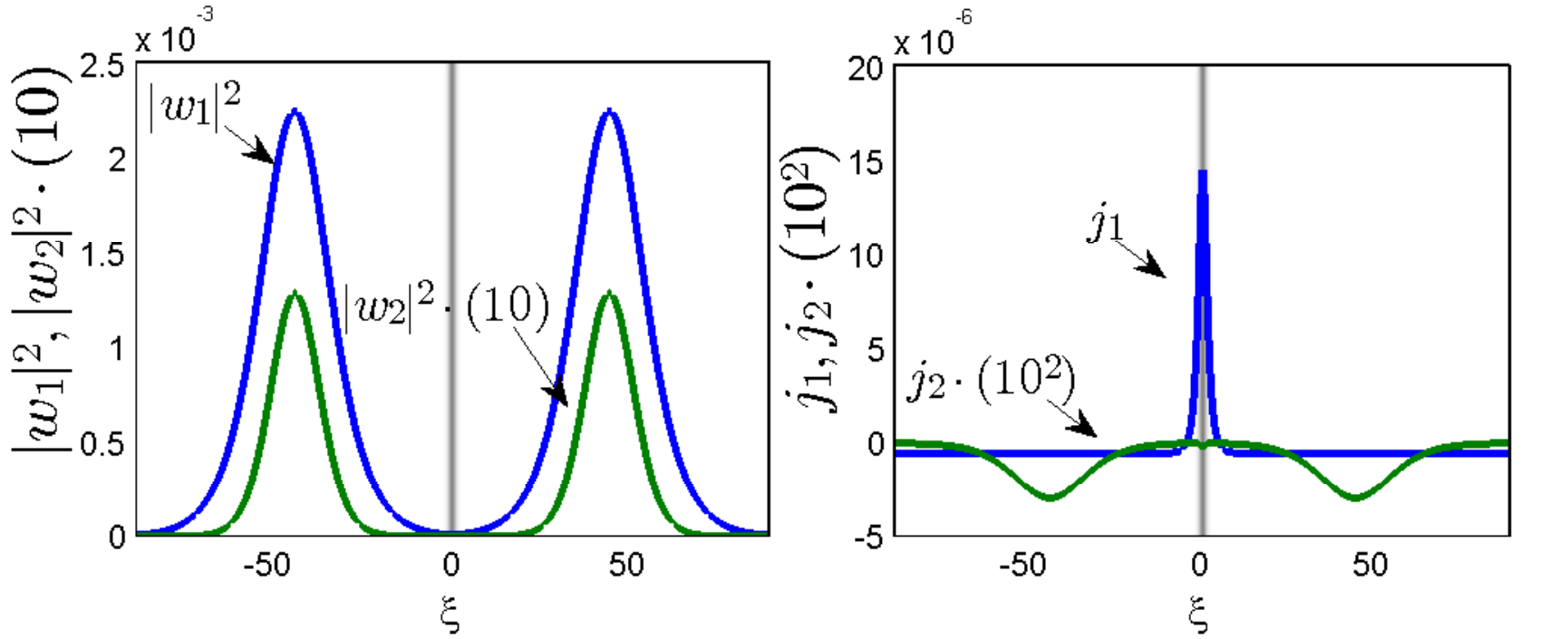}}
\end{center}
\caption{(Color online) Stable double peaked solution with $b=0.0057$ corresponding to the black circle in left panel of Fig.~\ref{fig:simb0}. Left panel shows the intensities $|w_n|^2$, right panel shows the currents $j_n$.  Shadowed domain show the localized impurity $V\text{sech}^{2}(\xi)$ (darker areas represent higher values of the real part of the localized potential). The parameters of the structure are $V=1/2$, $\alpha=1.3$ and $q=q_0$.}%
\label{fig:twopeak}
\end{figure}
 It is possible to see that two bifurcations occur at $b=b_{1,0}=b_{2,0}$ (See the dashed line in the insertion of Fig.~\ref{fig:V05qqbranch}). The branch that goes to the right  is a bifurcation of $b_{2,0}$ and the branch that goes to the left is a bifurcation of $b_{1,0}$. Both branches have a behaviour very similar to branches of case (i) and (ii).
\begin{figure}[h]
\begin{center}
\scalebox{0.55} {\includegraphics{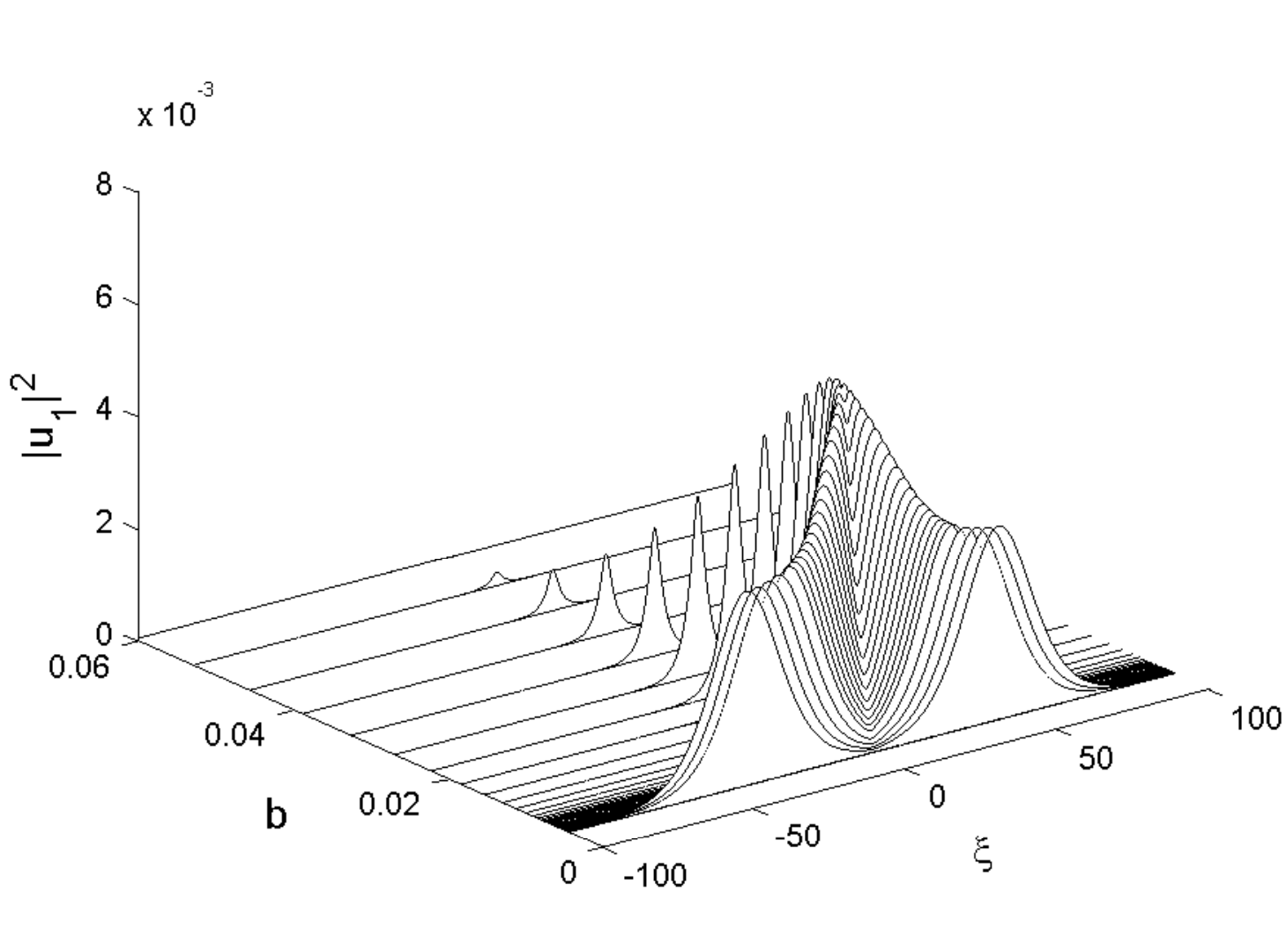}}
\end{center}
\caption{The intensity profiles $|w_{1}|^2$ of solutions,   pertaining to the fundamental branch of case (iii) bifurcating from $b_{1,0}=0.063$, at different $b$ illustrating the transition of a single-peaked profile into a double-peaked one. The local minimum occurs exactly at $\xi=0$. The parameters of the structure are $V=1/2$, $\alpha=1.3$ and $q=q_0$.}
\label{fig:transition}
\end{figure}

We observed in the numerical simulations, that there may be a collision of the fundamental branch with a non fundamental branch with two peaked solutions. The Fig.~\ref{fig:simb0} shows the corresponding bifurcation diagrams, whereas Fig.~\ref{fig:twopeak} illustrates the distribution of the intensities and the currents in a two-hump soliton solution. The intensities and the current $j_{2}$ are largely distributed far from the center of the potential, while the current $j_1$ is localized at the defect.

One can see in Fig.~\ref{fig:transition} that as $b$ decreases, $|w_1(0)|^2$ decreases at the same time that two emergent peaks become increasingly separated. The intensity $|w_2(0)|^2$ (not shown) decreases as well.

In respect to phase, we found all stable solutions that bifurcate from $b_{1,0}$ to satisfy $w_n(\xi)=\overline{w_n(-\xi)}$. This means that the peaks of the two-hump solution shown in Fig.~\ref{fig:transition} are out of phase.

\section{Stability and dynamics of localized solutions}

The stability was studied by direct numerical simulations  of the system  (\ref{final}) and within the framework of eigenvalue evaluation of the eigenvalue problem, obtained from perturbations of the form
\begin{figure}[h]
\begin{center}
\scalebox{0.6} {\includegraphics{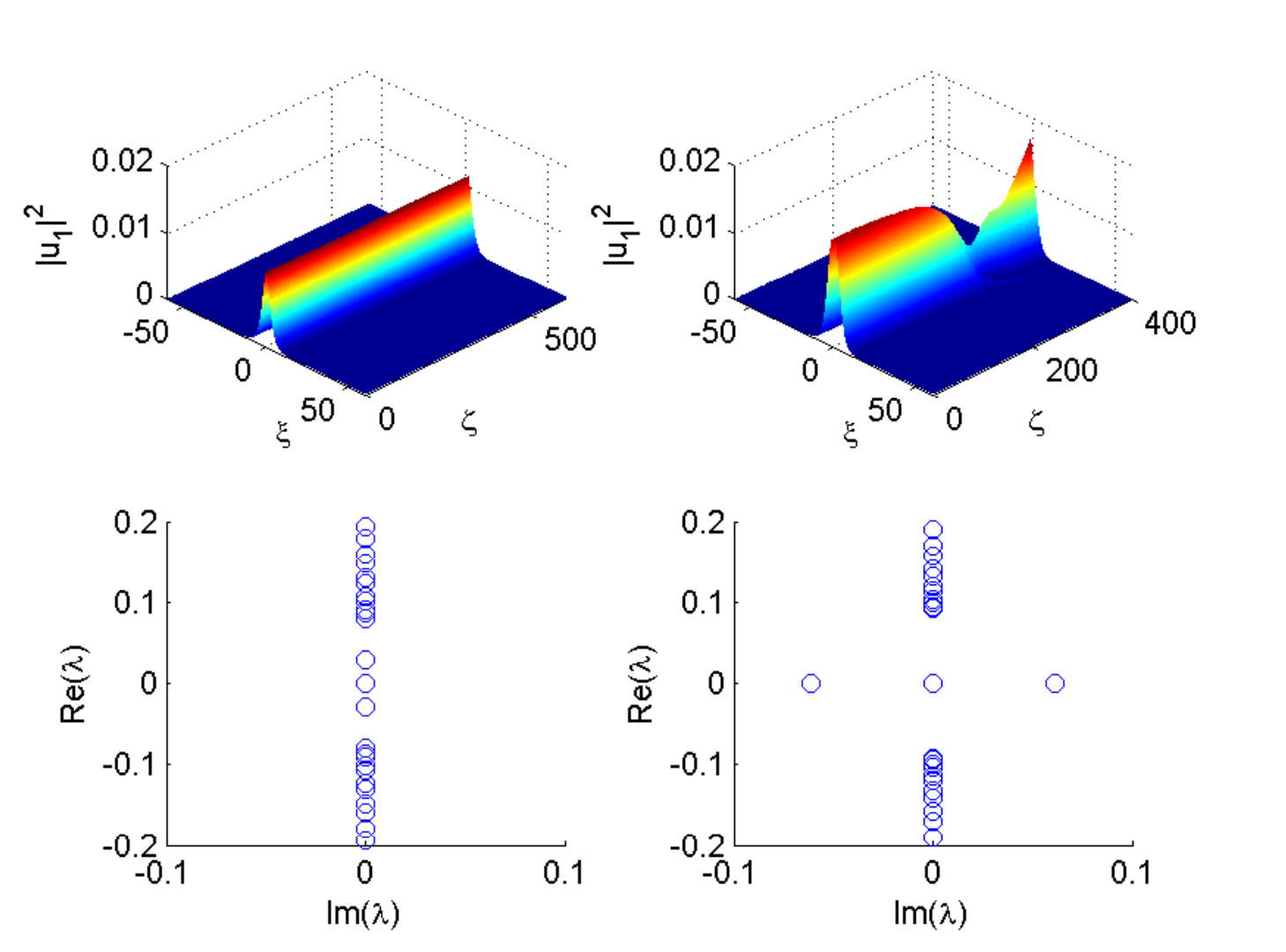}}
\end{center}
\caption{(Color online) The evolution of a stable solution with $b=0.076$ (left upper panel) and an unstable solution with $b=0.09$ (right upper panel) of the fundamental branch of case (i). The corresponding eigenvalues of the linear stability analysis are given in the lower panels. The parameters of the structure are $V=1/2$, $q=0$ and $\alpha=0.9$.}%
\label{fig:V05dinstabFF}%
\end{figure}
\begin{equation}
u_{n}(\xi,\zeta)=\left(  w_{n}(\xi)+p_{n+}(\xi)e^{-i\lambda\zeta
}+\overline{p}_{n-}(\xi)e^{i\overline{\lambda}\zeta}\right)  e^{in b\zeta
},
\end{equation}
 with $p_{n+}(\xi)$ and $p_{n-}(\xi)$ being small perturbations. The resulting eigenvalue problem is given by
\begin{equation}
\left(
\begin{array}
[c]{cccc}%
L_{2} & 2w_{1} & 0 & 0\\
2\bar{w}_{1} & L_{1,\alpha} & 0 & 2w_{2}\\
0 & 0 & -\bar{L}_{2} & -2\bar{w}_{1}\\
0 & -2\bar{w}_{2} & -2w_{1} & -\bar{L}_{1,\alpha}%
\end{array}
\right)  \left(
\begin{array}
[c]{c}%
p_{2+}\\
p_{1+}\\
p_{2-}\\
p_{1-}%
\end{array}
\right)  =\lambda\left(
\begin{array}
[c]{c}%
p_{2+}\\
p_{1+}\\
p_{2-}\\
p_{1-}
\end{array}
\right).  \label{linear}%
\end{equation}
Whenever an eigenvalue $\lambda$ with $\text{Im}(\lambda)>0$ occurs the solution is unstable.
\begin{figure}[h]
\begin{center}
\scalebox{0.65} {\includegraphics{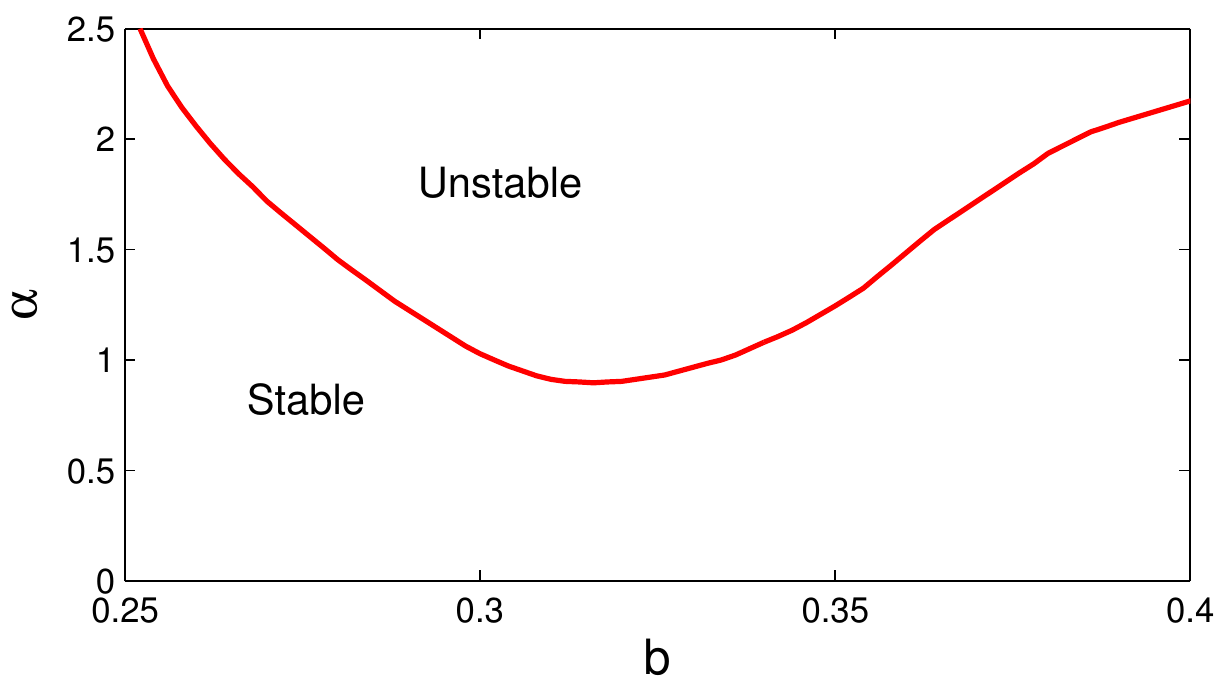}}
\end{center}
\caption{(Color online) Panel shows the maximal value of $\alpha$ for which a fundamental branch solution is stable as a function of $b$. The branch bifurcates from $b_{2,0}$. The parameters of the structure
are $V=1/2$ and $q=0$.}%
\label{fig:stab}%
\end{figure}

 Let us start the stability analysis with case (i). Then the branches have two stable regions, one close to $b_{1,0}$ and the other close to $b=0$ as is shown in Fig.~\ref{fig:branchsFF}. It was found numerically that only low amplitude solutions are stable. The instability is produced by pairs of purely imaginary eigenvalues $\lambda$ , see Fig.~\ref{fig:V05dinstabFF} showing the eigenvalues and the typical evolution of the soliton resulting in its rapid decay.

\begin{figure}[h]
\begin{center}
\scalebox{0.6} {\includegraphics{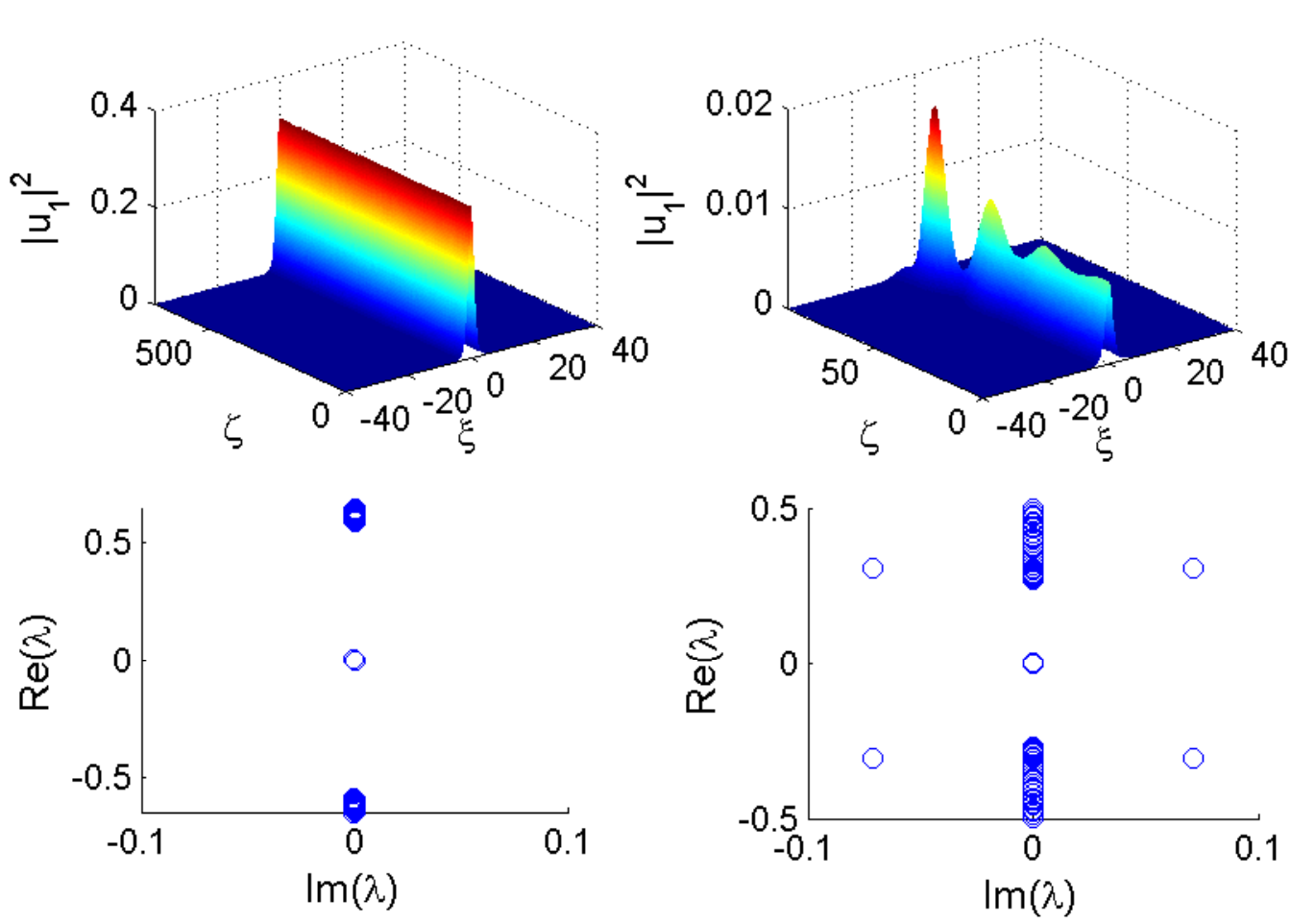}}
\end{center}
\caption{(Color online) The evolution of a stable solution with $b=0.6$ (left upper panel) and an unstable solution with $b=0.27$ (right upper panel) of the fundamental branch of case (ii) that bifurcates from $b_{2,0}$. Both solutions were perturbed by 10\% of amplitude random noise. The corresponding eigenvalues of the linear stability matrix are given in the lower panels. Both solutions are marked by black circles in Fig.~\ref{fig:branchSH}. The parameters of the structure are $V=1/2$, $q=0$ and $\alpha=2$.}%
\label{fig:stabSH}%
\end{figure}

 For case (ii), we investigate the stability and found that all the solutions are stable if $\alpha\lessapprox 0.9$, for $\alpha\gtrapprox 0.9$ some parts of the bifurcation curve become unstable, see Fig.~\ref{fig:stab}.
In  Fig.~\ref{fig:stab} one can see that the unstable part of the bifurcation curve becomes larger when $\alpha$ increases, but the stable solutions survive even for $\alpha \gg \alpha_{cr}$. 
The linear stability analysis shows that the instability arises from quartets of complex eigenvalues (See Fig.~\ref{fig:stabSH}).

\begin{figure}[h]
\begin{center}
\scalebox{0.6} {\includegraphics{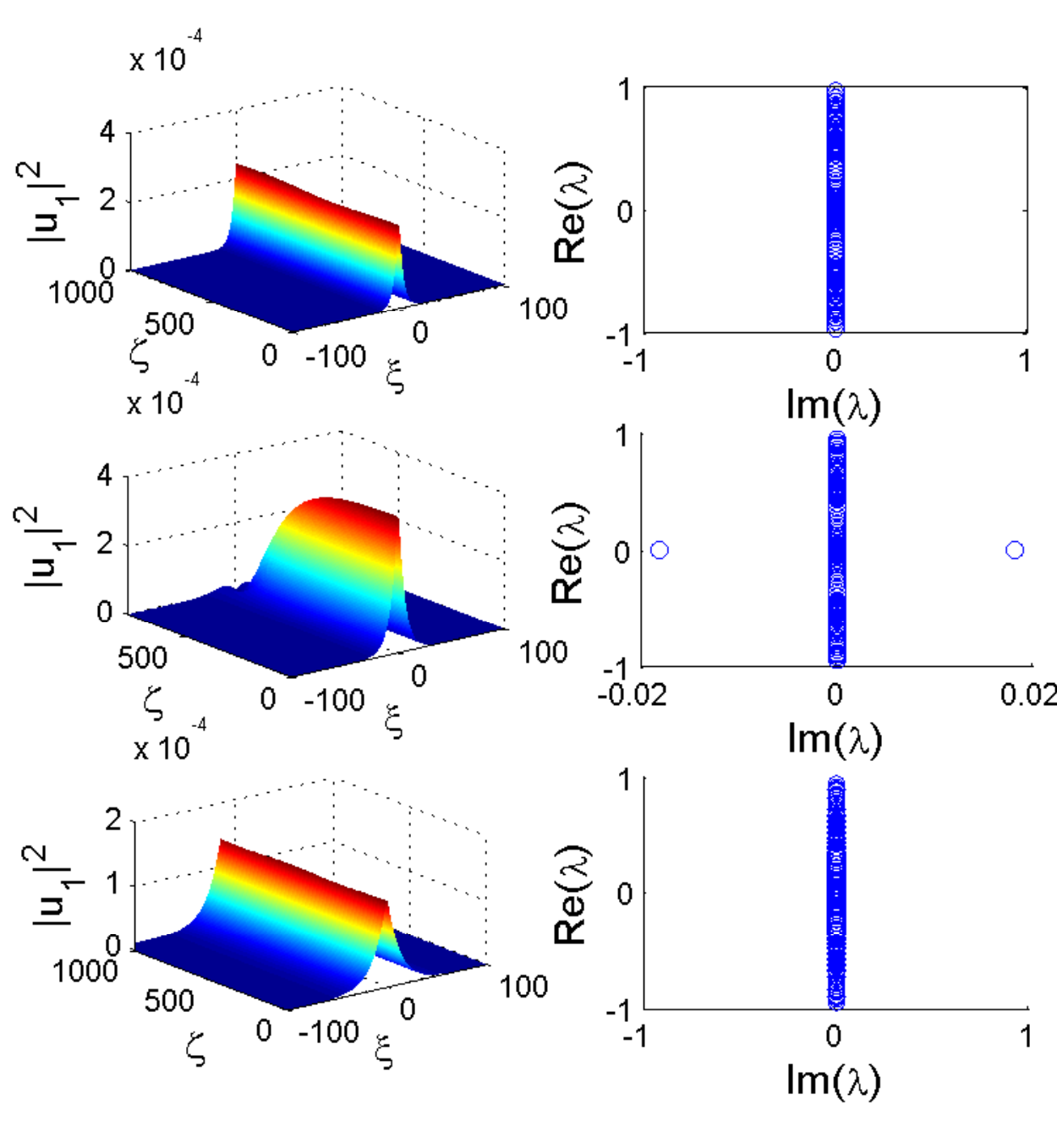}}
\end{center}
\caption{(Color online) Left panels: Propagation of 10\% of amplitude perturbations of solutions marked by black circles in lower panel of Fig.~\ref{fig:simb0} corresponding to case (iii). The upper left panel has $b=0.0172$, the midle left panel has $b=0.0072$ and lower left panel has $b=0.0017$. Right panels are the respective eigenvalues of the linear problem. Parameters of the structure are $V=1/2$, $\alpha=1.45$ and $q=q_0$.}%
\label{fig:dynstabSIMUL}%
\end{figure}

Finally we discuss the case (iii). We found that in respect to stability, the behaviour is similar to case (i) and (ii). For values $b>b_{1,0}$, i.e., bifurcations of $b_{2,0}$, the stability behaves like in case (ii), with the appearance of instability intervals that increase in length as $\alpha$ increases. The region $b<b_{1,0}$ has solutions that bifurcates from $b_{1,0}$. There is always a stable region adjacent to $b_{1,0}$ and also another stable region close to $b=0$. The instability, when observed, was due to a quartet of complex eigenvalues of the stability matrix contained in the region $b>b_{1,0}$ and two purely imaginary eigenvalues in the region contained in $b<b_{1,0}$ 
(See the middle panel of the Fig.~\ref{fig:dynstabSIMUL}). We observed that there are stable solutions in the region where the fundamental branch bifurcating from $b_{1,0}$ merges with a two-peaked branch. 


\section{Conclusion}
In summary, we show the existence of solitons in quadratic nonlinear media with localized $\PT$- symmetric modulations of linear refractive index.
The families of stable one and two-hump solitons are found. The properties of nonlinear modes bifurcating  from a linear limit of small fundamental harmonic field
are investigated.
 It is shown that the fundamental branch have a maximum in a power. This maximum is decreasing with the strength of of imaginary part of the refractive index $\alpha$. For the case when both harmonics are of the same order, the scenarios of bufurcations of different branches of solution on the propagation constant $b$  are investigated.It was shown that modes bifurcating from linear mode of the second harmonic can exist, even above $\PT$ symmetry breaking threshold. We found that the fundamental branch  bifurcating from the linear limit can undergo a secondary bifurcation colliding with a branch of two-hump soliton solutions.

For nonlinear modes no having linear limit i.e. $|u_2| \sim O(1)$, different branches of solutions in dependence on the parameters $b$ and the phase mismatch $q$ has been investigated. The stability intervals for different values of parameters $b$ and $\alpha$ are obtained. The examples of dynamics and excitations of solitons  by numerical simulations of full $\chi^{(2)}$ system of equations  with $\PT$ symmetric potential are  confirm theoretical predictions.

\acknowledgments
FCM akwnoledge the support of Alban. VVK and AVY acknowledge support of the FCT (Portugal) under the grants  PTDC/FIS/112624/2009, and PEst-OE/FIS/UI0618/2011.
FKA acknowledge support of the FAPESP(Brasil).


\begin{thebibliography}{99}

\bibitem{Bender1}
C. M. Bender and S. Boettcher, Phys.Rev.Lett. {\bf 80}, 5243 (1998);

\bibitem{Muga}
A. Ruschaupt, F. Delgado and J. G. Muga, J. Phys. A {\bf 38} L171 (2005).

\bibitem{Exp1}
A. Guo, G. J. Salamo, D.Duchesne,R. Morandotti, M. Volatier-Ravat, V. Aimez, G.A. Siviloglou, D.N. Christodoulides,
Phys. Rev. Lett., {\bf 103}, 093902 (2009); C. E. R\"uter, K.G.Makris, R. El-Ganainy,D.N. Christodoulides, M. Segev, and D. Kip, Nat. Phys. 6, 192 (2010);

\bibitem{Christodoulides1} Z. H. Musslimani, K. G. Makris, R. El-Ganainy, and D. N. Christodoulides
Phys. Rev. Lett. {\bf 100}, 030402, (2008).

%
%
%
%
%
%


\bibitem{Znojil}
M. Znojil, J.Phys. A Math. Gen. {\bf 33},  L61 (2000).


\bibitem{Ahmed}
Z. Ahmed, Phys. Lett. A {\bf 282}, 343-348 (2001).

\bibitem{Gaussian} Sumei Hu, Xuekai Ma, Daquan Lu, Zhenjun Yang, Yizhou Zheng, and Wei Hu    
Phys. Rev. A, {\bf 84}, 043818  (2011).



\bibitem{defoc} Zhiwei Shi, Xiujuan Jiang, Xing Zhu, and Huagang Li, Phys. Rev. A {\bf 84}, , 053855 (2011)

\bibitem{Mostafa_scattering} A. Mostafazadeh,
Phys. Rev. A, {\bf 80} 032711 (2009).

\bibitem{Kottos} O. Bendix, R. Fleischmann, T. Kottos, and B. Shapiro,
Phys. Rev. Lett. {\bf 103},  (2009)

\bibitem{AKOS} F. K. Abdullaev, V. V. Konotop, M. \"Ogren, and M. P. S\o rensen,
Opt. Lett., {\bf 36}, 4566 (2011).


\bibitem{double_well} H. Cartarius and G. Wunner  arXiv:1203.1885v1 (2012)

 \bibitem{PT_sol}
F. Kh. Abdullaev, V.V. Konotop, M. Salerno, and A. V. Yulin, Phys. Rev. E {\bf 82}, 056606 (2010);
F. Kh. Abdullaev, Y. V. Kartashov, V.V. Konotop, and D. A. Zezyulin, Phys. Rev. A {\bf 83}, 041805(R) (2011);
Y. He , X. Zhu, D. Mihalache, J. Liu and Z. Chen, Phys. Rev. A {\bf 85}, 013831 (2012);
S. Nixon, L. Ge, and J.Yang, Phys. Rev A {\bf 85}, 023822 (2012);
V. Achilleos, P. G. Kevrekidis,  D. J. Frantzeskakis, and R. Carretero-Gonzales,  Phys. Rev. A {\bf 86 }, 013808  (2012);
J. Zeng and Y. Lan, Phys.Rev. E {\bf 85}, 047601 (2012); S.V. Suchkov, S.V. Dmitriev, B. A. Malomed, and Yu. S. Kivshar, Phys. Rev. A {\bf 85}, 033825 (2012).


\bibitem{Clausen1}
 C.B. Clausen, J.P. Torres, and L. Torner, Phys. Lett. A {\bf 249}, 455 (1998).

\bibitem{Clausen2}
C.B. Clausen and L. Torner, Phys. Rev. Lett. {\bf 81}, 790 (1998).

\bibitem{Midya}
B. Midya, B. Roy, R. Roychoudhury, Phys. Lett. A {\bf 374} 2605-2607 (2010).

\bibitem{Landau}
D. Landau, E. M. Lifshitz, \textit{Quantum Mechanics: Non-Relativistic Theory, Volume {\bf 3}}, (Elsevier Science, MA, 1958)

\bibitem{Zezyulin}
D. A. Zezyulin and V.V. Konotop, Phys. Rev. Lett. {\bf 108}, 213906 (2012).

\bibitem{Tsoy}
E.N. Tsoy, S. Tadjimuratov, and F.Kh. Abdullaev, Opt. Commun. {\bf 285}, 3441 (2012).





\bibitem{Jacobi}
I.S. Gradshteyn, I.M. Ryzhik, Table of Integrals, Series and Products, 4th ed., Academic Press, New York, 1980, p. 838, (7.375.2).
\end{thebibliography}
\end{document}